# Perturbative study of the electroweak phase transition

Zoltán Fodor[*]

Deutsches Elektronen-Synchrotron, DESY, 22603 Hamburg, Germany

**Abstract**

The electroweak phase transition is studied at finite temperature. The effective action is given to higher orders, including wave function correction factors and the full $g^4, \lambda^2$ effective potential. An upper bound for the Higgs mass $m_H \approx 70\ GeV$ is concluded for the reliability of the perturbative approach. A gauge invariant treatment of the phase transition is presented.

## 1. Introduction

At high temperatures ($T$) the spontaneously broken electroweak symmetry is restored. Since the baryon-number violating processes are unsuppressed at high $T$, there is a possibility to understand the observed baryon asymmetry within the standard model [1]. However, a departure from thermal equilibrium, a sufficiently strong first order phase transition via bubble nucleation is needed.

In Sect. 2 the finite $T$ wave function corrections of the SU(2)-Higgs model to one-loop order [2] and the effective potential to order $g^4, \lambda^2$ will be studied [3]. This gives a range of Higgs boson masses ($m_H$) for which the derivative expansion of the effective action is reliable. Sect. 3 contains the gauge-invariant treatment of the finite $T$ electroweak effective potential [4]. This is of particular importance for comparison with lattice simulations [5], where the expectation value of $\Phi\Phi^\dagger$ is well suited to characterize the broken phase, and the corresponding effective potential has been evaluated [6].

## 2. The effective action at finite temperature

### 2.1. The wave function correction term

Consider the SU(2)-Higgs model at finite $T$, described by the lagrangian $\mathcal{L} = W^a_{\mu\nu} W^a_{\mu\nu}/4 + (D_\mu\Phi)^\dagger D_\mu\Phi + V_0(\varphi^2)$, where $V_0(\varphi^2) = m^2\varphi^2/2 + \mu\lambda\varphi^4/4$, $\varphi^2 = 2\Phi^\dagger\Phi$. $D_\mu$ and $W^a_{\mu\nu}$ are the covariant derivative and the Yang-Mills field strength, respectively. In this section Landau gauge is used and the effects of the three generations of fermions ($m_t = f_t v/\sqrt{2}$) have been included.

To get the effective action $\Gamma_\beta[\Phi]$ at finite temperature a systematic expansion is needed where in all propagators the tree-level masses are replaced by one-loop plasma masses to order $g^2$ and $\lambda$.

At one-loop order this improved perturbation theory yields the effective potential to order $g^3$, $\lambda^{3/2}$,

$$V_{eff}(\varphi^2, T) = \frac{1}{2}\left(\frac{3g^2}{16} + \frac{\lambda}{2} + \frac{1}{4}f_t^2\right)(T^2 - T_b^2)\varphi^2$$

$$+\frac{\lambda}{4}\varphi^4 - (3m_L^3 + 6m_T^3 + m_\varphi^3 + 3m_\chi^3)\frac{T}{12\pi}, \qquad (1)$$

which is equivalent to the result of the ring summation [7]. Here $m_L^2 = 11g^2T^2/6 + g^2\varphi^2/4$, $m_T^2 = g^2\varphi^2/4$, $m_\varphi^2 = (3g^2/16 + \lambda/2 + f_t^2/4)(T^2 - T_b^2) + 3\lambda\varphi^2$, $m_\chi^2 = (3g^2/16 + \lambda/2 + f_t^2/4)(T^2 - T_b^2) + \lambda\varphi^2$ and $T_b^2 = (16\lambda v^2)/(3g^2 + 8\lambda + 4f_t^2)$.

The strength of the electroweak phase transition is rather sensitive to the nonperturbative magnetic mass of the gauge bosons. In Landau gauge the one-loop gap equations yield $m_T = g^2T/(3\pi)$ at $\varphi = 0$. In order to estimate its effect we will replace [8] the previous definition of $m_T$ by $m_T^2 = \gamma^2 g^4 T^2/(9\pi^2) + g^2\varphi^2/4$ and compute sensitive quantities for different values of $\gamma$.

$V_{eff}$ of (1) has degenerate local minima at $\varphi = 0$ and $\varphi = \varphi_c > 0$ at a critical temperature

---

[*] On leave from Institute for Theoretical Physics, Eötvös University, Budapest, Hungary

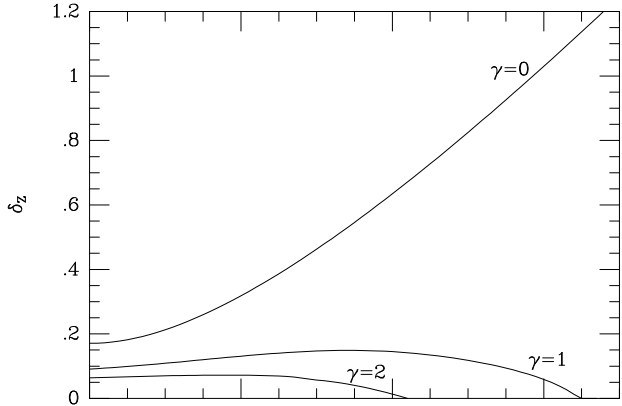

**Figure 1.** The one-loop wave function correction $\delta_Z$ as a function of $m_H$ for different values of $\gamma$.

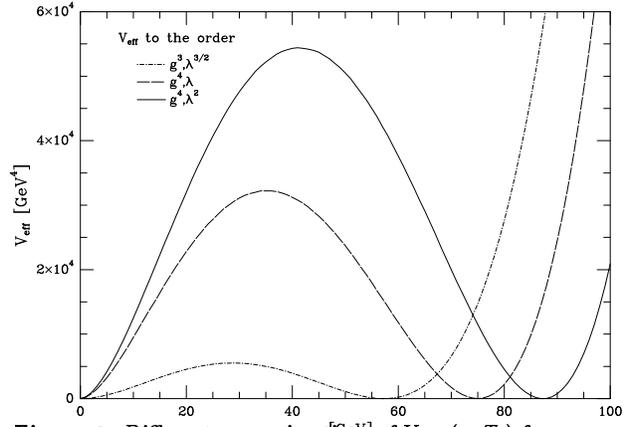

**Figure 2.** Different approximations of $V_{eff}(\varphi, T_c)$ for $m_H = 70~GeV$.

$T_c$. The evaluation of the transition rate requires knowledge of a stationary point of the free energy which interpolates between the two local minima. The effective action can be expanded in powers of derivatives, and for time-independent fields one has $T \cdot \Gamma_\beta[\Phi] = \int d^3x [V_{eff}(\varphi^2, T) + (\delta_{IJ} + Z_{IJ}(\Phi, T))\vec{\nabla}\varphi_I \vec{\nabla}\varphi_J/2 + \ldots]$. Using the inverse scalar propagator in the homogeneous scalar background field $\Phi$ one obtains on the one-loop level $Z_{IJ}(\Phi, T) = Z_\varphi(\varphi^2, T)P^\varphi_{IJ} + Z_\chi(\varphi^2, T)P^\chi_{IJ}$, where $Z_\varphi = T[\lambda \bar{m}^2(3/m_\varphi^3 + 1/m_T^3)/4 - 2g^2/(m_\chi + m_T) + g^2 m^2(1/m_L^3 + 10/m_T^3)/16]/(4\pi)$ and $Z_\chi = T[2\lambda \bar{m}^2/(m_\varphi + m_\chi)^3 - 2g^2/(m_\chi + m_T) - g^2/(m_\varphi + m_T)]/(6\pi)$ with $P^\varphi_{IJ} = \varphi_I\varphi_J/\varphi^2$, $P^\chi_{IJ} = \delta_{IJ} - \varphi_I\varphi_J/\varphi^2$, $\varphi^2 = \sum_{I=1}^4 \varphi_I\varphi_I$, $\bar{m}^2 = \lambda\varphi^2$ and $m^2 = g^2\varphi^2/4$.

Note, that despite the divergence of $Z_\varphi$ at $\varphi \sim 0$ the correction to the surface tension $\sigma = \int_0^{\varphi_c} d\varphi \sqrt{2(1 + Z_\varphi(\varphi^2, T_c))V_{eff}(\varphi^2, T_c)}$ is finite.

A measure for the size of the one-loop correction to the $Z$-factor is (Fig. 1) the ratio $\delta_Z = \int d^3x Z_{\bar\varphi}(\vec{\nabla}\bar\varphi)^2 / \int d^3x (\vec{\nabla}\bar\varphi)^2$, where $\bar\varphi$ is the saddle point solution at the nucleation temperature calculated from (1). For $\gamma = 0$ the perturbative expansion becomes unreliable at $m_H \sim 80$ GeV. The magnetic mass as an infrared cutoff ($\gamma = 1, 2$) could improve the convergence.

The above results are based on Ref. [2], where additional details can also be found.

### 2.2. The effective potential to order $g^4, \lambda^2$

The principal method [9] of the calculation is based on the Dyson-Schwinger equation for the derivative of the potential $\partial V/\partial\varphi$. The calculation has been done in an alternative way too, using the resummation [10] method of P. Arnold and O. Espinosa, who calculated the effective potential to order $g^4, \lambda$.

The full result of order $g^4, \lambda^2$ predicts a stronger first order phase transition than the lower order results. We plot the potential using different approximations for the $SU(2)$ Higgs-model at their respective $T_c$ (Fig. 2). The expectation value of the Higgs field does not change dramatically, but there is an order of magnitude difference between the heights of the barrier. No convergence of the perturbation series can be claimed for these parameters.

The surface tension, $\sigma = \int_0^{\varphi^+} d\varphi \sqrt{2V(\varphi, T_c)}$, may be seen as a measure of the strength of the phase transition (Fig. 3). The $g^4, \lambda^2$ result gives a radiatively induced quartic term and a better approximation of the temperature integrals, thus ensures that $\sigma$ does not grow for small $m_H$. For large $m_H$ the higher order result produces an increase in $\sigma$.

The full standard model calculation with zero temperature renormalization has also been done. The qualitative behaviour of the potential is essentially the same as for the $SU(2)$-Higgs model.

As it has been shown in the previous subsection wave function correction terms suggest that the perturbative approach becomes unreliable for large Higgs masses. In this subsection the analysis of the $g^4, \lambda^2$ contributions

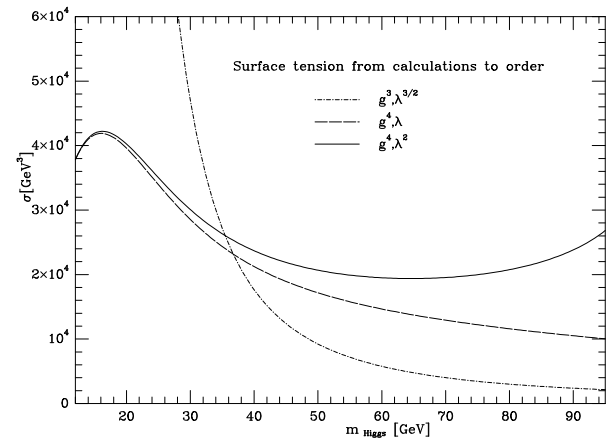

**Figure 3.** The surface tension calculated from the different potentials as a function of $m_H$.

to $V_{eff}$ led to a very similar conclusion.

The results of this subsection are from Ref. [3], where the details of the calculation and the full standard model analysis can also be found.

## 3. Gauge invariant treatment of the electroweak phase transition

In this section the effective potential for the composite field $\rho = 2\Phi^\dagger\Phi$ will be calculated [4]. For simplicity, the Higgs model in three dimensions is studied [11].

The SU(2)-Higgs model in three dimensions is described by the lagrangian $\mathcal{L} = W^a_{\mu\nu}W^a_{\mu\nu}/4 + (D_\mu\Phi)^\dagger D_\mu\Phi + V_0(\varphi^2)$, where $V_0(\varphi^2) = m^2\varphi^2/2 + \mu\lambda\varphi^4/4$, $\varphi^2 = 2\Phi^\dagger\Phi$. Here $W^a_{\mu\nu}$ is the ordinary field strength tensor and $D_\mu = \partial_\mu - i\sqrt{\mu}gW^a_\mu\tau^a/2$ is the covariant derivative; $\mu$ is the mass scale.

To obtain the effective potential for the field $\rho$, one evaluates the "free energy" in the presence of an external source $J$; $\exp[-\Omega W(J)] = \int DW D\Phi D\Phi^\dagger \exp[-S - \int d^3x 2\Phi^\dagger\Phi J]$, where $\Omega$ is the total volume. For constant $J$ one obtains $V(\rho)$ as a Legendre transform,

$$\partial W(J)/\partial J = \rho \; , \quad V(\rho) = W(J(\rho)) - \rho J \; . \quad (2)$$

$W(J)$ can be calculated in the semiclassical or loopwise expansion. The equation for a spatially constant stationary point, $(m^2 + 2\mu\lambda\Phi^\dagger_c\Phi_c + 2J)\Phi_c = 0$, has two solutions, which correspond to the symmetric and the broken phase, respectively, $\Phi_s = 0$ and $\Phi_b = \Phi_0[-(m^2 + 2J)/(2\mu\lambda)]^{1/2}$, where $|\Phi_0| = 1$. The determinants of fluctuations around the two stationary points depend on the masses of vector bosons, Higgs ($\varphi$) and Goldstone ($\chi$) bosons. In the broken phase ($\Phi_c = \Phi_b$) one has, in any covariant gauge, $m^2_W = -g^2(m^2 + 2J)/(4\lambda)$, $m^2_\varphi = -2(m^2 + 2J)$, $m^2_\chi = 0$, whereas in the symmetric phase ($\Phi_c = \Phi_s$) the masses are given by $m^2_W = 0$, $m^2_\varphi = m^2_\chi = m^2 + 2J$.

The one-loop $W(J)$ in covariant gauge is

$$W_1(J) = \frac{1}{2}\int \frac{d^3k}{(2\pi)^3} \left(6\ln(k^2 + m^2_W) + \ln(k^2 + m^2_\varphi)\right.$$
$$\left. + 3\ln(k^4 + k^2 m^2_\chi + \alpha m^2_W m^2_\chi) - 6\ln k^2\right) \; , \quad (3)$$

where $\alpha$ is the gauge parameter. This expression is gauge independent and the same result for $W_1(J)$ is obtained in $R_\xi$-gauge.

Subtracting linear divergencies by means of dimensional regularization and performing the Legendre transformation according to eq. (2) gives

$$V(\rho) = V_b(\rho)\Theta(\rho) + V_s(\rho)\Theta(-\rho) \; , \quad (4)$$

where $V_b(\rho) = m^2\rho/2 + \mu\lambda\rho^2/4 - [6(\mu g^2\rho/4)^{3/2} + (2\mu\lambda\rho)^{3/2}]/(12\pi)$ and $V_s(\rho) = m^2\rho/2 - \pi^2\rho^3/6$. Here the couplings depend on the renormalization scale, i.e., $g = g(\mu)$, $\lambda = \lambda(\mu)$, $m^2 = m^2(\mu)$.

The Landau gauge effective potential for the field $\Phi$

$$V_{LG}(\varphi^2) = \frac{m^2\varphi^2}{2} + \frac{\mu\lambda\varphi^4}{4} - \frac{6m^3_W + m^3_\varphi + 3m^3_\chi}{12\pi} \; , \quad (5)$$

with $m^2_W = \mu g^2\varphi^2/4$, $m^2_\varphi = m^2 + 3\mu\lambda\varphi^2$, $m^2_\chi = m^2 + \mu\lambda\varphi^2$.

Comparing the two potentials (4) and (5) the first difference is the range of the fields. For (5) one has $0 \leq \varphi^2 < \infty$, whereas for (4) the field varies in the range $-\infty < \rho < \infty$. In (5) the symmetric phase is represented by the point $\varphi = 0$, whereas in (4) by the half-axis $\rho \leq 0$. At small values of $\rho$ the potential increases very steeply. The second important difference is that the non-analytic terms of the gauge invariant potential do not depend on $m^2$. Hence, this potential can also be used for $m^2 < 0$, where the symmetric phase is unstable.

We have performed the above calculation for the SU(2)-Higgs model at finite $T$. The gauge invariant potential [4] is $V(\rho, T) = V_s(\rho, T)\Theta(\rho) + V_b(\rho, T)\Theta(-\rho)$, where $V_b(\rho, T) = m^2(T)\rho/2 + \lambda\rho^2/4 - T[6(g^2\rho/4)^{3/2} + (2\lambda\rho)^{3/2}]/(12\pi)$ and $V_s(\rho, T) = m^2(T)\rho/2 - \pi^2\rho^3/(6T^2)$. Contrary to the conventional potential (neglect the fermions in $V_{eff}(\varphi^2, T)$ of eq. 1), $V(\rho, T)$ is valid at temperatures above and below $T_b$.

We have evaluated several observables for the conventional and for the gauge invariant effective potentials. The critical temperatures are different, but very similar. For Higgs masses between 30 GeV and 120 GeV the ratio $(T_c - T_b)/T_b$ differs by at most 40%. The latent heat differs by about 70% at $m_H = 120$ GeV.